\documentclass[twocolumn,showpacs,floatfix,superscriptaddress,amsmath,amssymb,prl]{revtex4}
\usepackage{mathrsfs}
\usepackage{txfonts}
\usepackage{amssymb}
\usepackage{graphicx}
\usepackage{hyperref}
\usepackage{ulem}
 \usepackage{overpic}
 \usepackage{psfrag}
\usepackage{tabularx}
\usepackage{array}
\usepackage{placeins}
\makeatletter
\renewcommand\subsection{\@startsection
{subsection}{2}{0mm}
 {-\baselineskip}
 {0.5\baselineskip}
{\FloatBarrier\normalfont\Large\bfseries}}
\makeatother
\newcommand{\be}{\begin{equation}}
\newcommand{\ee}{\end{equation}}

\newcommand{\PreserveBackslash}[1]{\let\temp=\\#1\let\\=\temp}

\begin{document}
\title{Quantum phase transitions and bifurcations: reduced fidelity as a phase transition
indicator for quantum lattice many-body systems}

\author{Jin-Hua Liu}
\affiliation{Centre for Modern Physics and Department of Physics,
Chongqing University, Chongqing 400044, The People's Republic of
China}
\author{Qian-Qian Shi}
\affiliation{Centre for Modern Physics and Department of Physics,
Chongqing University, Chongqing 400044, The People's Republic of
China}
\author{Jian-Hui Zhao}
\affiliation{Centre for Modern Physics and Department of Physics,
Chongqing University, Chongqing 400044, The People's Republic of
China}
\author{Huan-Qiang Zhou}
\affiliation{Centre for Modern Physics and Department of Physics,
Chongqing University, Chongqing 400044, The People's Republic of
China}

\begin{abstract}
We establish an intriguing connection between quantum phase
transitions and bifurcations in the reduced fidelity between two
different reduced density matrices for quantum lattice many-body
systems with symmetry-breaking orders.   Our finding is based on the
observation that,  in the conventional Landau-Ginzburg-Wilson
paradigm, a quantum system undergoing a phase transition is
characterized in terms of spontaneous symmetry breaking that is
captured by a local order parameter, which in turn results in an
essential change of the reduced density matrix in the
symmetry-broken phase. Two quantum systems on an infinite lattice in
one spatial dimension, i.e., quantum Ising model in a transverse
magnetic field and quantum spin 1/2 XYX model in an external
magnetic field, are considered in the context of the tensor network
algorithm based on the matrix product state representation.

\end{abstract}

\pacs{03.67.-a, 03.65.Ud, 03.67.Hk}

\maketitle

{\it Introduction.} Recently we have witnessed a growing interest in
the study of quantum many-body systems in the context of the
fidelity approach to quantum phase transitions (QPTs)~\cite{b1,b2}
with symmetry-breaking/topological
orders~\cite{zanardi,b6,zhou,zhou1,b7,b8,b9,b3,fidelity,reducedfidelity}.
In Refs.~\cite{b6,zhou,zhou1,b7}, it has been argued that the ground
state fidelity per site may be used to detect QPTs.  Since the
argument is solely based on the basic Postulate of Quantum Mechanics
on quantum measurements, it is expected that this approach is
applicable to quantum lattice systems in any spatial dimensions,
regardless of what type of internal order is present in quantum
many-body states. In fact, it has been confirmed that the ground
state fidelity per site is able to describe QPTs arising from an SSB
~\cite{b6,zhou,zhou1,b7,b3}, the Kosterlitz-Thouless
transition~\cite{b9} and topological QPTs in the Kitaev
model~\cite{b8}. Remarkably, the ground state fidelity per site may
be computed in terms of the newly-developed tensor network (TN)
algorithms, such as the matrix product states (MPS)~\cite{b10,b11,
b12} in one spatial dimension, and the tensor product states
(TPS)~\cite{tps}, or equivalently, the projected entangled-pair
states (PEPS)~\cite{peps}, in two and higher spatial dimensions.

In the conventional Landau-Ginzburg-Wilson paradigm, an SSB, which
occurs when a system possesses a certain symmetry whereas the ground
state wave functions do not preserve it~\cite{b4, b5}, is quantified
in terms of a local order parameter. An intriguing feature of local
order parameters for quantum systems with symmetry-breaking orders
is bifurcations they exhibit at critical points. Remarkably, such a
bifurcation also manifests itself in the ground state fidelity per
site~\cite{b3}. The advantage of the latter over local order
parameters lies in the fact that the ground state fidelity per site
is \textit{universal} in the sense that it is not model-dependent,
in contrast to model-dependent order parameters in characterizing
QPTs in quantum lattice many-body systems.

The investigation that has been carried out so far mainly focuses on
the ground state wave functions for quantum lattice systems. Thus
the fidelity is equivalent to the overlap between two different
ground states, which are pure states. However, as first discussed in
Ref.~\cite{zhou1} (see also\;\cite{reducedfidelity}), the fidelity
defined for two different mixed states is also useful to locate
phase transition points. The mixed states are  described by the
reduced density matrices arising from tracing out the degrees of
freedom in the environment surrounding a sub-system. Indeed, for a
sub-system of a composite quantum system, a reduced density matrix
is a basic notion that is indispensable for the analysis of a
composite quantum system. In fact, the presence of a local order
parameter makes a difference in the reduced density matrices in the
symmetric and symmetry-broken phases. From this one may anticipate
that the (reduced) fidelity between two different reduced density
matrices for quantum lattice many-body systems with
symmetry-breaking orders should capture bifurcations arising from an
SSB.

In this paper, we attempt to address this problem. We shall
investigate two quantum models on an infinite lattice in one spatial
dimension, i.e., quantum Ising model in a transverse magnetic field
and spin 1/2 XYX model in an external magnetic field.  Both systems
possess a discrete symmetry group $Z_{2}$, but the ground states
break the symmetry. The SSB is reflected as a bifurcation~\cite{b3}
in the ground state reduced fidelity for both systems. Our result
demonstrate that one may identify a phase transition point as a
bifurcation point in the ground state reduced fidelity between two
different reduced density matrices for quantum lattice many-body
systems with symmetry-breaking orders~\cite{twod}.

{\it The models.}   The first model we consider in this paper is the
quantum Ising model in a transverse magnetic field in an
infinite-size lattice in one spatial dimension. The Hamiltonian
takes the form,
\begin{equation}\label {fit1}
   H = - \sum_{i=-\infty}^{\infty} (S_{x}^{[i]}S_{x}^{[i+1]}+\lambda
   S_{z}^{[i]}),
 \end{equation}
where $S_{\alpha}^{[i]} \; (\alpha = x, z)$ are the spin 1/2 Pauli
operators at site $i$, and $\lambda$ is the transverse magnetic
field. The model is invariant under the symmetry operation:
$S_{x}^{[i]} \rightarrow -S_{x}^{[i]}$ and  $S_{z}^{[i]} \rightarrow
S_{z}^{[i]}$ for all sites, which yields the $Z_{2}$ symmetry.  As
is well known, the system undergoes a second order QPT at the
critical field $\lambda_{c}  = 1$~\cite{cris}.

The second model is the spin 1/2 XYX model in an external magnetic
field. The Hamiltonian can be written as
\begin{equation}\label {fit2}
   H = \sum_{i=-\infty}^{\infty} (S_{x}^{[i]} S_{x}^{[i+1]} + \Delta_{y}S_{y}^{[i]} S_{y}^{[i+1]} + S_{z}^{[i]} S_{z}^{[i+1]} +
   hS_{z}^{[i]}),
 \end{equation}
where $S_{\alpha}^{[i]} \; (\alpha = x,y,z)$ are the Pauli spin
operators at site $i$, $\Delta_{y}$ is a parameter describing the
anisotropy in the internal space, and $h$ is the external magnetic
field. This model also possesses a $Z_{2}$ symmetry, with the
symmetry operation: $S_{x}^{[i]} \rightarrow -S_{x}^{[i]}$,
$S_{y}^{[i]} \rightarrow -S_{y}^{[i]}$ and $S_{z}^{[i]} \rightarrow
S_{z}^{[i]}$ for all sites. Below we shall choose $\Delta_{y} =
0.25$. In this case, the critical magnetic field is $h_{c} \sim
3.210(6)$~\cite{mc}.

{\it The reduced density matrix.} We are now in a position to
clarify the difference of the reduced density matrices in the
symmetric and symmetry-broken phases for both quantum Ising model in
a transverse magnetic field and quantum spin 1/2 XYX model in an
external magnetic field. The analysis will be carried out for both
the one-site and two-site reduced density matrices, respectively.
For the quantum Ising model in a transverse magnetic field, the
one-site reduced density matrix in the $Z_{2}$ symmetric phase takes
the form,
\begin{equation}\label {fitoneising}
   \rho_{ising} = \frac{1}{2} + 2\langle S_{z}\rangle S_{z},
 \end{equation}
where $\langle S_{z}\rangle$ is the expectation value of $S_z$ in
the ground state in the $Z_2$ symmetric phase, whereas the two-site
reduced density matrix in the $Z_{2}$ symmetric phase is,
\begin{eqnarray}\label {fitwoising}
   \rho_{ising} &=& \frac{1}{4}I + 4\alpha_{1} S_{x} \otimes S_{x} + 4\alpha_{2} S_{z} \otimes S_{z} \notag\\
       && + \alpha_{3} I \otimes S_{z} + \alpha_{4} S_{z} \otimes I.
\end{eqnarray}
Here, $\alpha_{1}=\langle S_{x} \otimes S_{x} \rangle$,
$\alpha_{2}=\langle S_{z} \otimes S_{z} \rangle$,
$\alpha_{3}=\langle I \otimes S_{z} \rangle$, and
$\alpha_{4}=\langle S_{z} \otimes I \rangle$, with $I$ being the
identity matrix.

In the $Z_{2}$ symmetry-broken phase, the presence of the nonzero
local order parameter $\langle S_{x}\rangle$ implies that the
one-site reduced density matrix takes the form,
\begin{equation}\label {fitanoneising}
   \rho_{ising} = \frac{1}{2} + 2\langle S_{x}\rangle S_{x} + 2\langle S_{z}\rangle
   S_{z}.
 \end{equation}
Such a violation of the symmetry is also reflected in  the two-site
reduced density matrix:
\begin{eqnarray}\label {fitantwoising}
   \rho_{ising} &=& \frac{1}{4}I + 4\alpha_{1} S_{x} \otimes S_{x} + 4\alpha_{2} S_{z} \otimes S_{z} \notag\\
       && + \alpha_{3} I \otimes S_{z} + \alpha_{4} S_{z} \otimes I + + 4\alpha_{5} S_{x} \otimes S_{z} \notag\\
       && + 4\alpha_{6} S_{z} \otimes S_{x} + \alpha_{7} I \otimes S_{x} + \alpha_{8} S_{x} \otimes
       I,
\end{eqnarray}
with $\alpha_{5}=\langle S_{x} \otimes S_{z} \rangle$,
$\alpha_{6}=\langle S_{z} \otimes S_{x} \rangle$,
$\alpha_{7}=\langle I \otimes S_{x} \rangle$, and
$\alpha_{8}=\langle S_{x} \otimes I \rangle$.

For the quantum spin 1/2 XYX model in an external magnetic field,
the one-site reduced density matrix in the $Z_{2}$ symmetric phase
takes the form,
\begin{equation}\label {fitsonex}
   \rho_{XYX} = \frac{1}{2} + 2\langle S_{z}\rangle S_{z},
 \end{equation}
while the two-site  reduced density matrix is
\begin{eqnarray}\label {fitsonx}
   \rho_{XYX} &=& \frac{1}{4}I + 4\beta_{1} S_{x} \otimes S_{x} + 4\beta_{2} S_{y} \otimes S_{y} \notag\\
       && +  4\beta_{3} S_{z} \otimes S_{z} + \beta_{4} I \otimes S_{z} +  \beta_{5} S_{z} \otimes I,
\end{eqnarray}
with $\beta_{1}=\langle S_{x} \otimes S_{x}\rangle$,
$\beta_{2}=\langle S_{y} \otimes S_{y}\rangle$, $\beta_{3}=\langle
S_{z} \otimes S_{z}\rangle$, $\beta_{4}=\langle I \otimes
S_{z}\rangle$, and $\beta_{5}=\langle S_{z} \otimes I\rangle$.

In the symmetry-broken phase, the one-site reduced density matrix
becomes
\begin{equation}\label {fitaonex}
   \rho_{XYX} = \frac{1}{2} +2\langle S_{x}\rangle S_{x} + 2\langle S_{z}\rangle S_{z},
 \end{equation}
whereas the two-site reduced density matrix is
\begin{eqnarray}\label {fitswox}
   \rho_{XYX} &=& \frac{1}{4}I + 4\beta_{1} S_{x} \otimes S_{x} + 4\beta_{2} S_{y} \otimes S_{y} +  4\beta_{3} S_{z} \otimes S_{z} \notag\\
       && + \beta_{4} I \otimes S_{z} +  \beta_{5} S_{z} \otimes I + 4\beta_{6} S_{x} \otimes S_{z} \\ \notag   && +4\beta_{7} S_{z} \otimes S_{x} + \beta_{8} I \otimes S_{x}
      + \beta_{9} S_{x} \otimes I,
\end{eqnarray}
with $\beta_{6}=\langle S_{x} \otimes S_{z}\rangle$,
$\beta_{7}=\langle S_{z} \otimes S_{x}\rangle$, $\beta_{8}=\langle I
\otimes S_{x}\rangle$, and $\beta_{9}=\langle S_{x} \otimes I
\rangle$.

{\it Bifurcations in the reduced fidelity between two reduced
density matrices and quantum phase transitions.} The reduced
fidelity measures the distance between two quantum mixed states.
Specifically, for two reduced density matrices $\rho_{\lambda}$ and
$\rho_{\lambda^{'}}$, the reduced fidelity
$F(\rho_{\lambda},\rho_{\lambda^{'}})$ is defined to be
\begin{equation}\label {fit}
   F(\rho_{\lambda},\rho_{\lambda^{'}})=tr\sqrt{ \rho_{\lambda}^{1/2}  \rho_{\lambda^{'}}
   \rho_{\lambda}^{1/2}}.
 \end{equation}
Here, $\rho_{\lambda}$ and $\rho_{\lambda^{'}}$ are the reduced
density matrices corresponding to two different values, $\lambda$
and $\lambda^{'}$, of the control parameter $\lambda$. Notice that
the reduced fidelity $F(\rho_{\lambda},\rho_{\lambda^{'}})$ is a
function of $\lambda$ and $\lambda^{'}$, which satisfies the
following properties: (i) normalization
$F(\rho_{\lambda},\rho_{\lambda})=1$; (ii) symmetry
$F(\rho_{\lambda},\rho_{\lambda^{'}})=F(\rho_{\lambda^{'}},\rho_{\lambda})$;
(iii) range 0$ \leq F(\rho_{\lambda},\rho_{\lambda^{'}})\leq $1.

The connection between a bifurcation point in the ground state
partial fidelity between two reduced density matrices and a critical
point for quantum lattice systems undergoing QTPs with
symmetry-breaking order may be established in the following way. In
the conventional Landau-Ginzburg-Wilson paradigm, a quantum system
undergoing a QPT is characterized in terms of an SSB that is
captured by a local order parameter, which in turn results in an
essential change of the reduced density matrix in the symmetry
broken phase, as seen above. More precisely, consider a quantum
system on an infinite lattice in one spacial dimension, with $Z_{2}$
as a symmetry group~\cite{footnote}. Suppose the $Z_{2}$ symmetry is
spontaneously broken when the control parameter $\lambda$ crosses a
critical point $\lambda =\lambda_c$, then the system undergoes a QPT
with a nonzero local order parameter in the symmetry-broken phase.
Such a local order parameter is defined in a local area $\Omega$ on
a lattice, and it is \textit{not} invariant under the action of the
symmetry operation generating the symmetry group $Z_2$, thus
yielding degenerate ground states in the symmetry-broken phase.
Therefore, the reduced fidelity
$F(\rho_{\lambda},\rho_{\lambda^{'}})$, with the reference state
$\rho_{\lambda^{'}}$ in the symmetry-broken phase, yields
\textit{two} different values, which correspond to two distinct
values of the local order parameter arising from two degenerate
ground states, if $\rho_{\lambda}$ is in the symmetry-broken phase,
and \textit{one} value, if $\rho_{\lambda}$ is in the symmetric
phase. That is, a bifurcation point occurs in the ground state
reduced fidelity between two reduced density matrices
$F(\rho_{\lambda},\rho_{\lambda^{'}})$, with the critical point
being the bifurcation point. However, if we choose the reference
state $\rho_{\lambda^{'}}$ in the symmetric phase, then no such a
bifurcation occurs, due to the invariance of the reduced density
matrix $\rho_{\lambda^{'}}$ under the action of the symmetry
operation. In addition, the same argument applies to the reduced
fidelity defined on any local area $\Omega'$ with $\Omega$ as a
subset~\cite{monotonic}.

\begin{figure}
\begin{overpic}[width=85mm,totalheight=60mm]{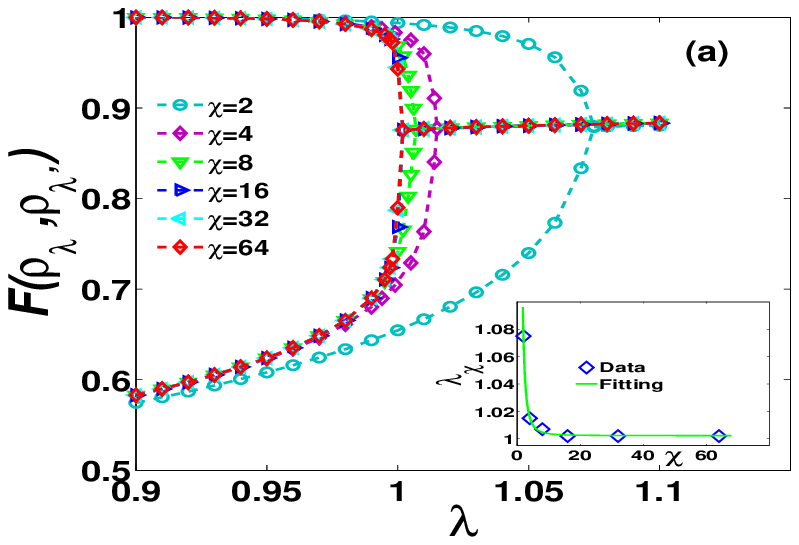}
\end{overpic}
\begin{overpic}[width=85mm,totalheight=60mm]{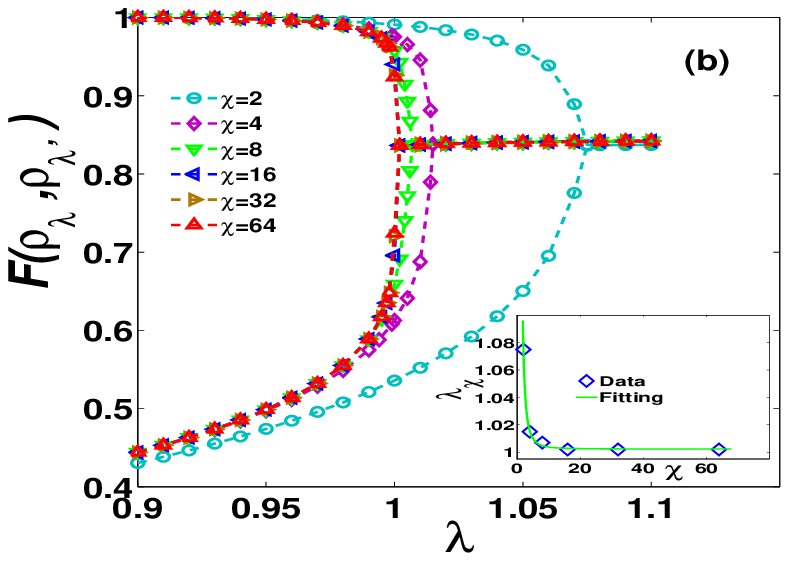}
\end{overpic}
\setlength{\abovecaptionskip}{0pt} \caption{(color online) (a) The
ground state one-site reduced fidelity
$F(\rho_{\lambda},\rho_{\lambda^{'}})$ for the quantum Ising model
in a transverse magnetic field.  We choose $\rho_{\lambda^{'}}$
(with $\lambda^{'} = 0.9$) as the reference state, which is in the
$Z_2$ symmetry-broken phase.   The pseudo phase transition point
$\lambda_{\chi}$ occurs as a bifurcation point. When we enlarge the
truncation dimension $\chi$, the pseudo phase transition point
$\lambda_{\chi}$ is getting closer to the exact value. Inset: the
critical point $\lambda_{c}$ is determined from an extrapolation of
the pseudo phase transition point $\lambda_{\chi}$  with respect to
the truncation dimension $\chi$. Here, the fitting function is
$\lambda_{\chi} = \lambda_{c} + a\chi^{-b}$, with $\lambda_{c} =
1.00233$,\;$ a = 0.39373$ and $b = 2.43904$.  (b) The ground state
two-site reduced fidelity, $F(\rho_{\lambda},\rho_{\lambda^{'}})$,
for the quantum Ising model in a transverse magnetic field.  The
same reference state $\rho_{\lambda^{'}}$ \; ($\lambda^{'} = 0.9$)
has been chosen as in the case of the one-site reduced fidelity.
Both insets indicate the same pseudo critical points for the ground
state one-site and two-site reduced density matrices for the system.
This is expected due to the fact that they are resulted from the
same set of the ground states.} \label{FIG1}
\end{figure}

\begin{figure}
 \begin{overpic}[width=85mm,totalheight=60mm]{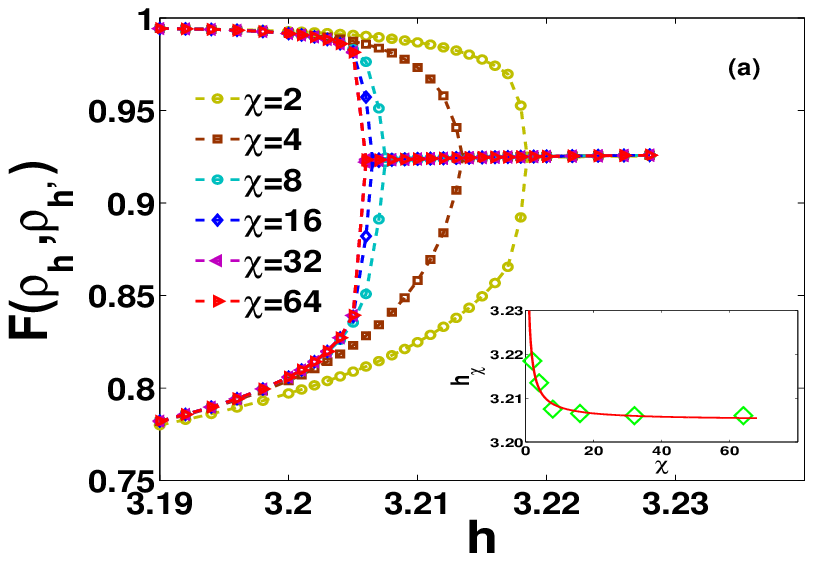}
\end{overpic}
\begin{overpic}[width=85mm,totalheight=60mm]{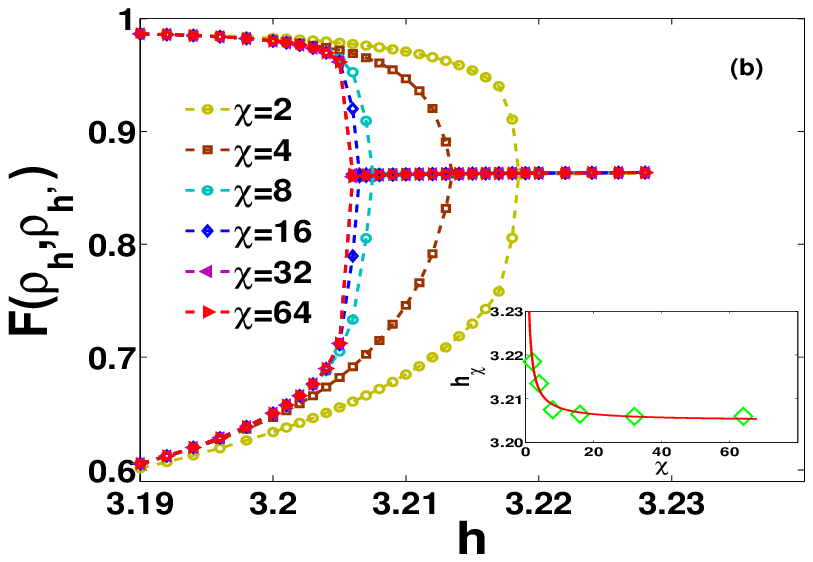}
\end{overpic}
\setlength{\abovecaptionskip}{0pt} \caption{(color online) (a) The
ground state one-site reduced fidelity $F(\rho_{h},\rho_{h^{'}})$
for the quantum spin 1/2 XYX model in an external magnetic field.
Here, the external magnetic field strength $h$ is the control
parameter. We choose $\rho_{h^{'}}$\; ( with $h^{'} = 3.05$) as the
reference state, which is in the symmetry-broken phase.   The pseudo
phase transition point $h_{\chi}$ occurs  as a bifurcation point.
For a larger value of $\chi$, the pseudo phase transition point
$h_{\chi}$ is getting closer to the known value 3.2049. Inset: the
critical point $h_{c}$ is determined from an extrapolation of the
pseudo phase transition point $h_{\chi}$ with respect to the
truncation dimension $\chi$. The fitting function is $h_{\chi} =
h_{c} + a\chi^{-b}$, with $h_{c} = 3.2049$, $a = 0.0267$ and $b =
0.9426$. (b) The  ground state two-site reduced fidelity
$F(\rho_{h},\rho_{h^{'}})$ for the quantum spin 1/2 XYX model. We
still choose $\rho_{h^{'}} \;( h^{'} = 3.05)$ as the reference
state. The two-site reduced fidelity exhibits the similar behavior
to that of the one-site reduced fidelity. Both insets display the
same pseudo critical points for the one-site  and two-site reduced
fidelity for the system.} \label{FIG2}
\end{figure}

{\it The results.} In the context of the TN algorithm initiated by
Vidal~\cite{vidal}, the problem to find the system's ground state
wave functions amounts to computing the imaginary time evolution for
a given initial state $|\Psi (0)\rangle $: $ |\Psi (\tau)\rangle =
\exp(- H \tau) |\Psi (0)\rangle / |\exp(- H \tau) |\Psi (0)\rangle
|$. An efficient way to achieve this task is to exploit the
Suzuki-Trotter decomposition~\cite{su}, which allows us to reduce
the imaginary time evolution operation to a product of two-site
evolution operators acting on sites $i$ and $i+1$: $U(i,i+1) =
\exp(-h^{[i,i+1)]} \delta \tau )$, $\delta \tau <<1$. In addition,
any wave function admits an MPS representation in a canonical form:
attached to each site is a three-index tensor $\Gamma^s_{A\;lr}$ or
$\Gamma^s_{B\;lr}$, and to each bond a diagonal singular value
matrix $\lambda_A$ or $\lambda_B$, depending on the evenness and
oddness of the $i$-th site and the $i$-th bond. Here, $s$ is a
physical index, $s=1,\cdots,d$, with $d$ being the dimension of the
local Hilbert space, and $l$ and $r$ denote the bond indices,
$l,r=1,\cdots, \chi$, with $\chi$ being the truncation dimension.
The action of a two-site gate $U(i,i+1)$ may be absorbed by
performing a singular value decomposition, thus resulting in the
update of the MPS representation. Repeating this procedure until the
ground state energy converges, one may generate the ground state
wave functions in the MPS representation. We emphasize that, in
practice, we adjust the truncation dimension $\chi$ to identify a
critical point from bifurcation points in the reduced fidelity
$F(\rho_{\lambda},\rho_{\lambda^{'}})$. In fact, it saves a lot of
the computational resources to perform simulations for relatively
small values of $\chi$. Usually, a shift in the bifurcation points
occurs due to the finiteness of $\chi$. Therefore, it is necessary
to perform an extrapolation with respect to $\chi$ to locate the
critical point.

In Fig.\ref{FIG1}(a), we plot the ground state reduced fidelity
$F(\rho_{\lambda},\rho_{\lambda^{'}})$ between the one-site reduced
density matrices for the quantum Ising model in a transverse field
with the field strength $\lambda$ as the control parameter. Here, we
choose $\rho_{\lambda^{'}}\;(\lambda^{'}=0.9)$ as a reference state,
which breaks the $Z_{2}$ symmetry.  The one-site reduced fidelity
can distinguish two degenerate ground states with a bifurcation
point as a pseudo phase transition point $\lambda_{\chi}$~\cite{b3}.
When the parameter $\lambda$ is tuned beyond such a pseudo
transition point, two degenerate ground states vanish, implying that
the system undergoes a phase transition. One observes that, with
$\chi$ increasing, the pseudo phase transition point
$\lambda_{\chi}$ moves toward the exact value 1.  Performing an
extrapolation of $\lambda_{\chi}$ with respect to $\chi$, we  get
$\lambda_{c}=1.00233$. In Fig.\ref{FIG1}(b), we show the two-site
reduced fidelity for the quantum Ising model in a transverse
magnetic field. The same reference state is selected as in the case
of the one-site reduced fidelity. We observe that a bifurcation also
occurs in the two-site partial fidelity.  Indeed, it yields the same
pseudo phase transition points $\lambda_{\chi}$, and thus the same
critical point $\lambda_{c}$.

In Fig. \ref{FIG2}(a), the ground state one-site reduced fidelity
$F(\rho_{h},\rho_{h^{'}})$ for the quantum spin 1/2 XYX model in an
external magnetic field $h$ is plotted, with the magnetic field
strength $h$ as the control parameter. We choose $\rho_{h^{'}}$,
with $h^{'} = 3.05$, in the $Z_{2}$ symmetry-broken phase, as the
reference state. The reduced fidelity $F(\rho_{h},\rho_{h^{'}})$
distinguishes two degenerate ground states in the symmetry-broken
phase.  The bifurcation point $h_{\chi}$ resulted from the one-site
reduced fidelity is the pseudo phase transition point, which is
quite close to the  known critical value $h_{c} \sim
3.210(6)$~\cite{mc}, if the truncation dimension  $\chi$ is large
enough.  Performing an extrapolation with respect to $\chi$ yields
the critical point $h_{c} = 3.2049$. In Fig. \ref{FIG2}(b), we plot
the two-site reduced fidelity for the quantum XYX model. Here, the
same reference state as in the case of the one-site partial fidelity
has been chosen. The two-site reduced fidelity
$F(\rho_{h},\rho_{h^{'}})$  is also able to detect the $Z_{2}$ SSB.
Indeed, it again yields the same bifurcation points $h_{\chi}$, and
thus the same critical point $h_{c}$.

{\it Summary.} We have established an intriguing connection between
QPTs and bifurcations in the reduced fidelity between two different
reduced density matrices for quantum lattice many-body systems with
symmetry-breaking orders.   Our work is based on the newly-developed
TN algorithms, which produce degenerate ground states in the
symmetry-broken phase for quantum lattice systems under QPTs arising
from an SSB.  Two quantum systems on an infinite lattice in one
spatial dimension, i.e., quantum Ising model in a transverse
magnetic field and quantum spin 1/2 XYX model in an external
magnetic field, have been investigated in the context of the TN
algorithm based on the MPS representation.

{\it Acknowledgements.} This work is supported in part by the
National Natural Science Foundation of China (Grant Nos: 10774197
and 10874252), the Natural Science Foundation of Chongqing (Grant
No: CSTC, 2008BC2023).



\begin{thebibliography}{10}
\bibitem{twod} Although our examples are restricted to quantum
systems on an infinite lattice in one spatial dimension, the
argument is valid for quantum systems in two and higher dimensions.
In the latter case, the TN algorithm based on PEPS~\cite{ipeps} is
necessary.
\bibitem{footnote}  The extension of our argument to any other group is
straightforward. It also applies to quantum lattice systems in any
spatial dimensions.
\bibitem{monotonic} Moreover, one may establish the monotonicities
of the reduced fidelity $F(\rho_{\lambda},\rho_{\lambda^{'}})$ with
$\lambda$ for a fixed $\lambda'$, along a similar line to the ground
state fidelity per site~\cite{b6}, which may be used to identify
stable fixed points.
\end{thebibliography}
\end{document}